\documentclass[aps,preprint,amssymb,superscriptaddress]{revtex4-1}
\usepackage{floatrow}
\newfloatcommand{capbtabbox}{table}[\FBwidth]
\usepackage{graphicx}
\usepackage{dcolumn}
\usepackage{bm}
\usepackage{natbib}
\usepackage{CJK}
\usepackage[utf8]{inputenc}
\usepackage[T1]{fontenc}
\usepackage{mathptmx}
\usepackage{gensymb}
\usepackage{graphicx}
\usepackage{epstopdf}
\usepackage{dcolumn}
\usepackage{bm}
\usepackage{amsmath}
\usepackage{mathtools}
\usepackage{relsize}
\usepackage{multirow}
\usepackage[colorlinks]{hyperref}
\begin{document}


\title{Investigating relative population partitioning in the presence of B$_\text{C}^-$ defects in diamond}

\author{Rodrick Kuate Defo}
\email{rkuatede@syr.edu}
\affiliation{Department of Electrical Engineering and Computer Science, Syracuse University, Syracuse, New York 13244, USA}
\affiliation{Department of Physics, Syracuse University, Syracuse, New York 13244, USA}
\author{Steven L. Richardson} 
\affiliation{John A. Paulson School of Engineering and Applied Sciences, Harvard University, Cambridge, MA 02138, USA}
\affiliation{Department of Electrical and Computer Engineering, Howard University, Washington, DC 20059}
\date{\today}

\begin{abstract}
Leveraging our earlier formalism for charge-state decay [see R. Kuate Defo \textit{et al.}, \textbf{108} 235208 (2023)], we investigate relative population partitioning in the presence of singly negatively charged substitutional boron impurities (B$^-_{\text{C}}$) in diamond samples, where these diamond samples also contain substitutional nitrogen (N$_{\text{C}}$) impurities and have ether-terminated surfaces. Effectively, given any two defects, one donor and one acceptor, such that their charge-transition levels, their relative populations, and the temperature are known, the timescale for charge transfer between the two defects is completely determined. We use this insight and experimental measurements for the timescale for charge transfer between NV$^-$ centers and N$_{\text{C}}$ hybridized with the ether-terminated surface of diamond, as well as the temperature and the relative populations for the NV$^-$ centers, [see Z. Yuan et al., Phys. Rev. Res. \textbf{2}, 033263 (2020)] to argue that an additional defect must be modulating the relative populations of N$_{\text{C}}$ hybridized with the ether-terminated surface of diamond. The shifts in the local Fermi level for N$_{\text{C}}$ hybridized with the ether-terminated surface of diamond are consistent with the additional defect being a substitutional boron impurity (B$_{\text{C}}$). Ultimately, our results allow for the determination of local relative concentrations of ionized defects.
\end{abstract}

\maketitle

\section{Introduction}
Interfacing with NV$^-$ centers in diamond is typically done through the extraction of fluorescence~\cite{Mizuochi2012,Lohrmann2011}. We had previously described how by analogy with the spin-state dependent fluorescence of the NV$^-$ center, the fact that the NV$^0$ is visible under electroluminescence, but not the NV$^-$~\cite{Mizuochi2012}, could effectively be used to implement charge-state dependent fluorescence~\cite{kuate_investigating_2024}. This charge-state dependent fluorescence has a potential application in the realization of charge-state-based spin readout for NV centers~\cite{Giri,Giri2023charge,Shields2015efficient,Ariyaratne2018nanoscale,Bar-Gill2013solid,zhang2025achieving,Jayakumar2018spin,Jaskula2019improved,Hopper2018amplified,Rovny2022nanoscale} in bulk diamonds~\cite{Jayakumar2018spin, Jaskula2019improved} and in nanodiamonds~\cite{Hopper2018amplified}. A persistent issue is the existence of surface and near-surface electron traps~\cite{Dhomkar2018charge,Giri,Stacey2019evidence,Giri2023charge,Kuate2023theor,Mittiga2018imaging}, which detrimentally affect charge-state stability~\cite{Bluvstein2019identifying,Giri2023charge,Giri}. We and others~\cite{Kuate2023theor,Kuate2023charge,Kuate2026erratum,Mittiga2018imaging,kuate_investigating_2024} had identified that the leading-order contribution to the charge-state decay characteristics of the NV$^-$ center is from the abundant substitutional nitrogen (N$_{\text{C}}$) defects present in typical diamond samples, though others have argued instead that such charge-state stability behavior could be explained by the presence of interstitial-related defects~\cite{Li2026depth}. 

We aim in this work to demonstrate that our ordering of contributions is borne out in the analysis of another aspect of the Yuan \textit{et al.} 2020~\cite{yuan2020charge} experiment, namely their investigation of NV$^-$ charge-state decay characteristics in samples persistently contaminated with boron. We begin by presenting the details of our computations in Section~\ref{sec:methods}. We next proceed to outline our theoretical formalism in Section~\ref{sec:theorapproach}. We then provide a discussion of our results in Section~\ref{sec:disc}, and, finally, we conclude in Section~\ref{sec:conc}.

\section{Computational Methods}
\label{sec:methods}
The calculations we performed in this work were for the substitutional boron (B$_{\text{C}}$) impurity in diamond. For these calculations, we employed the Vienna Ab Initio Simulation Package (VASP)~\cite{Kresse1993ab,Kresse1996efficient,Kuate2021theor,Kresse1999from}. We used the HSE06 functional with the original parameters~\cite{Heyd,Krukau}.  We terminated our calculations when the forces dropped below a threshold of 10$^{-4}$~eV/\AA~in the atomic-position relaxations. The wavefunctions were expanded in a planewave basis with a cutoff energy of 500~eV. The size of the supercell was 512 atoms (a $4\times4\times4$ multiple of the conventional unit cell of diamond). Reciprocal-space integrations were performed at the $\Gamma$ point. 

\section{Theoretical Approach}
\label{sec:theorapproach}
We had previously described an iterative approach~\cite{kuate_investigating_2024,bahulikar_theoretical_2025} whereby first ionized defects with levels closest to one another would exchange charge and then ionized defects or collections of ionized defects with defect levels farther apart would exchange charge, once sufficient time had elapsed. The timescales for the exchanges of charge are completely determined by the charge-state transition levels of the defects involved in the charge transfer as well as by the relative populations of the ionized charge states of the defects involved in the charge transfer and the temperature~\cite{Kuate2023charge,Kuate2026erratum}. In the Yuan \textit{et al.} experiment~\cite{yuan2020charge}, the relative populations for the NV$^-$ centers as well as the timescales for charge transfer had been specified. Given previous determination that the leading order contribution to the shift in the optical transition levels of the NV$^-$ should come from N$_{\text{C}}$~\cite{Kuate2023theor,Mittiga2018imaging}, we consider charge exchange for NV$^-$ centers with N$_{\text{C}}^+$, which have been hybridized with the ether-terminated surface typically found in oxygen-terminated diamond samples~\cite{Kuate2023charge,kuate_investigating_2024}. We find that the timescale for charge transfer in the Yuan \textit{et al.} experiment~\cite{yuan2020charge} is consistent with an additional negative shift in the local Fermi level associated with the N$_{\text{C}}^+$ defect hybridized with the ether-terminated surface of diamond, compared to the shift resulting only from hybridization of the N$_{\text{C}}$ with the ether-terminated surface of diamond. This additional negative shift is consistent with the presence of B$_{\text{C}}^-$ in the diamond sample.

We had written that the timescale of charge transfer is given by~\cite{Kuate2026erratum,kuate_investigating_2024,Kuate2023charge},
\begin{equation}
\tau \approx \frac{1}{\pi}\int_0^\pi d\theta\left(\frac{1}{\Delta t}\frac{1}{\left(\exp\left(\left(E_{\text{F}_\text{D}}-E_{\text{F}_\text{A}}-k_BT\cos^2\left(\theta\right)\text{sgn}(\cos(\theta))\right)/\left(k_BT\right)\right)+1\right)}\right)^{-1},
\end{equation}   

\begin{figure}[ht!] 
\centering
\includegraphics[width=0.7\textwidth]{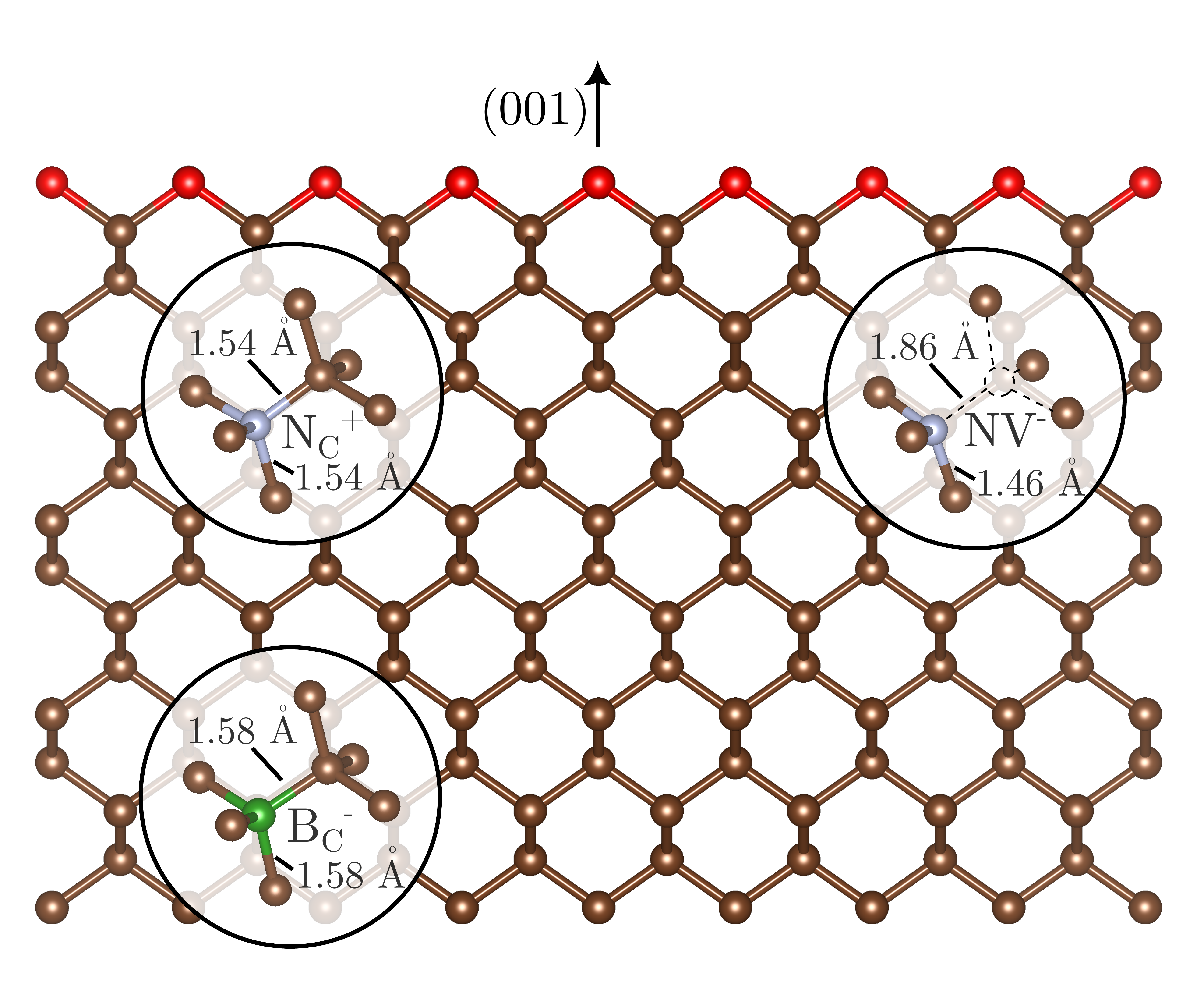}
\caption{Arrangement of charged N$_{\text{C}}$, NV, and B$_{\text{C}}$ defects with (111) orientation in unrelaxed (001) oriented diamond with an ether-terminated surface. The O atoms are in red, the C atoms are in brown, the N atoms are in purple, and the B atom is in green.}
\label{fig:quadrupole}
\end{figure}
\noindent where $\Delta t = \int_{r_0}^{0}dr\left(\frac{dr}{dt}\right)^{-1}$ for some separation $r_0$ between NV and N$_\text{C}$ defects, $T$ is temperature, $t$ is time, and $k_B$ is Boltzmann's constant. We also use the velocity defined through the equation~\cite{Kuate2026erratum},
\begin{equation}
\frac{dr}{dt} =  \lim_{\alpha \rightarrow 1^+} \pm\sqrt{2/m^*\left(E_{\text{F}_\text{D}}-E_{\text{F}_\text{A}}-\left(\frac{r}{r_0}\right)^\alpha\left(E_{\text{F}_\text{D}}-E_{\text{F}_\text{A}}-k_BT\cos^2(\theta)\right)\right)},
\end{equation}
where $m^*$ is the effective mass of the electron. Additionally, $E_{\text{F}_\text{D}}-E_{\text{F}_\text{A}}$ is taken to represent a difference in local Fermi levels between a donor and an acceptor and is explicitly computed as $E_{\text{F}_\text{D}}-E_{\text{F}_\text{A}} = e\left(\phi(\mathbf{r}_{\text{X}^\text{q}}) - \phi(\mathbf{r}_{\text{Y}^{\text{q}'}})\right) = e\left(\phi(\mathbf{r}_{\text{N}^{+}_\text{C}}) - \phi(\mathbf{r}_{\text{NV}^{-}})\right)$ with $\phi(\mathbf{r}_{\text{X}^\text{q}})$ and $\phi(\mathbf{r}_{\text{Y}^{\text{q}'}})$ being the values of the electromagnetic potential at the locations of defects X = N$_\text{C}$ (the donor) and Y = NV (the acceptor), $\text{q} = +1$ and $\text{q}' = -1$ being charge states, and $e$ being the elementary charge. With two defects, the problem could be approximated as a one-dimensional problem. In the presence of B$_{\text{C}}$, we will continue to approximate the problem as a one-dimensional problem since we can argue that there are sufficiently many B$_{\text{C}}$ in the sample (approximately 4\% boron presence in the surface monolayer as measured by x-ray photoelectron spectroscopy~\cite{yuan2020charge}) that the distance between the NV$^-$ and the N$_\text{C}^+$ can be viewed as large compared to the distance from the N$_\text{C}^+$ to the nearest B$_{\text{C}}$ that could be polarized by the N$_\text{C}^+$. The potentials are defined relative to the neutral system as~\cite{Kuate2023charge,Kuate2023theor},
\begin{align}
\phi\left(\mathbf{r}_{\text{X}^{\text{q}}}\right) &= -\frac{1}{e\text{q}}\left(\Delta H_f\left(\text{X}^{\text{q}},\left\{\mu_i^\text{X}\right\},0\right)-\Delta H_f\left(\text{X}^{\text{0}},\left\{\mu_i^\text{X}\right\},0\right)\right)\\
&= -\frac{1}{e\text{q}}\left[\left(E_{\text{def}}\left(\text{X}^{\text{q}}\right)+E_{\text{corr}}\left(\text{X}^{\text{q}}\right)\right)-E_{\text{def}}\left(\text{X}^{\text{0}}\right)\right],
\end{align}
where $\Delta H_f\left(\text{X}^{\text{q}},\left\{\mu_i^\text{X}\right\},0\right)$ is the formation energy associated with the species X$^\text{q}$ with charge q, $E_{\text{def}}\left(\text{X}^{\text{q}}\right)$ is the total energy of the supercell with the $\text{X}^{\text{q}}$ species, and $E_{\text{corr}}\left(\text{X}^{\text{q}}\right)$ is a charge-state dependent electrostatic correction energy to account for interaction with periodic images~\cite{Vinichenko,Freysoldt2011electrostatic,Freysoldt2009fully,Kumagai,Komsa2013finite,Walsh2021}. 

\section{Results and Discussion}
\label{sec:disc}
\begin{figure}[ht!] 
\centering
\includegraphics[width=0.475\textwidth]{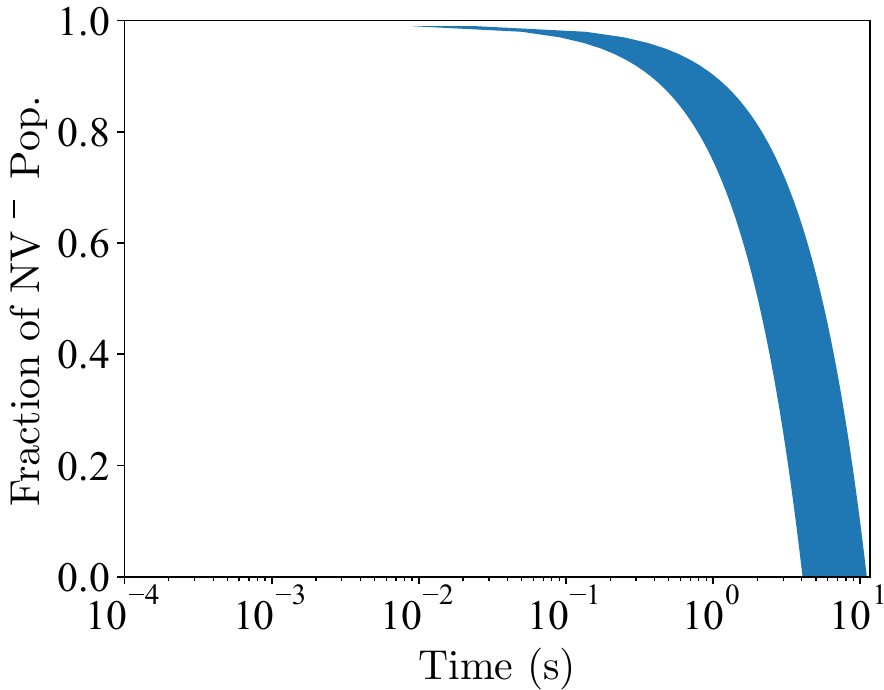}
\caption{Charge-transition timescales calculated for bounding relative populations of $p^{\text{NV}^-}_{\text{min}} \approx 0.19$ and $p^{\text{NV}^-}_{\text{max}} \approx 0.40$ using $T = 300$~K. The defect concentration employed for N$_{\text{C}}$ is $n_{\text{N}_{\text{C}}} = \frac{10^9~\text{cm}^{-2}}{d_{\text{max}}}$. Here $d_\text{max} = (3.5~\text{nm/keV})\cdot E_{\text{imp}}$~\cite{Broadway2018spat}, where $E_{\text{imp}} = 1.5~$keV~\cite{yuan2020charge}. The effect of the presence of B$_{\text{C}}$ has not been included. The effect of the ether-terminated surface, however, has been included (see Refs.~\cite{Kuate2026erratum,kuate_investigating_2024,Kuate2023charge}).}
\label{fig:NVm1_decay_woB_wB}
\end{figure}

Based on our calculations, referenced to the valence band maximum (VBM), we find a local potential $\phi(\mathbf{r}_{{\text{B}^-_\text{C}}}) \approx 0.3$~eV/$e$, which is in good agreement with the acceptor level of 0.37~eV typically found in the literature~\cite{Klimova2026engineering}. We had previously calculated $\phi(\mathbf{r}_{{\text{NV}^-}}) \approx 2.8$~eV/$e$ and $\phi(\mathbf{r}_{{\text{N}^+_{\text{C}}}}) \approx 3.6$~eV/$e$~\cite{Kuate2023charge,Kuate2023theor,kuate_investigating_2024} (see also Ref.~\cite{Weber2010quantum}). In all cases, the VBM is located at VBM $\approx 9.4$~eV in the absolute units of VASP and the conduction band minimum (CBM) is located at CBM $\approx 14.8$~eV in VASP's absolute units. Furthermore, for an ether-terminated surface, the defect level 3.4~eV had been provided by experiment and by other theory~\cite{Sque2006structure,Broadway2018spat}. From the Yuan \textit{et al.} experiment~\cite{yuan2020charge}, we employ the bounding relative populations at 1~s $p^{\text{NV}^-}_{\text{min}} \approx 0.19$ and $p^{\text{NV}^-}_{\text{max}} \approx 0.40$. The result of na\"ively employing those relative populations without accounting otherwise for the presence of B$_{\text{C}}$ is displayed in Fig. \ref{fig:NVm1_decay_woB_wB}. 

We had previously defined the relative population correction for the NV$^-$ through the equation~\cite{Kuate2023charge},
\begin{equation}
\label{eq:shiftpop}
\Delta E_{\text{F}} = k_BT\ln\left(\frac{\left(1-p_0\right)p}{p_0\left(1-p\right)}\right),
\end{equation}
where $\Delta E_{\text{F}} = E_{\text{F}} - \epsilon^{\text{NV}}\left(0/-\right)$ is the shift in the local Fermi level due to the relative population of NV$^-$ defects with $\epsilon^{\text{NV}}\left(0/-\right)$ being the 0 to $-1$ charge-transition level for the NV center, $p$ is the relative population of NV$^-$ defects, and $p_0 = 0.5$ is a reference relative population of NV$^-$ defects. For the N$_{\text{C}}$ and ether-terminated diamond surface complex, which we'll label as $\eta$, following the same definition of the relative population correction as for the NV$^-$, we have $\Delta E_{\text{F}}' = E_{\text{F}}' - \epsilon^{\eta}\left(0/+\right)$. The shifts $\Delta E_{\text{F}}'$ that allow the bounding charge-state decay curves each to end at 1~s, corresponding to the time at which the relative populations for the NV$^-$ centers were determined, as alluded to above, are $\Delta E_{\text{F}}' \approx -0.063$~eV for $p^{\text{NV}^-}_{\text{min}} \approx 0.19$ and $\Delta E_{\text{F}}' \approx -0.037$~eV for $p^{\text{NV}^-}_{\text{max}} \approx 0.40$. The negative shift in the local Fermi level in both cases is consistent with the proposal that the N$_{\text{C}}$ and ether-terminated diamond surface are hybridizing with a B$_{\text{C}}$ defect. We can add the shifts to the shift associated with N$_{\text{C}}$ hybridization with the ether-terminated diamond surface to determine the relative populations for N$_{\text{C}}^+$ through Eq. (\ref{eq:shiftpop}). Using $T = 300$~K, we find that when $p^{\text{NV}^-}_{\text{min}} \approx 0.19$, $p^{\text{N}_{\text{C}}^+}_{\text{min}} \approx 0.00087$. Similarly, when $p^{\text{NV}^-}_{\text{max}} \approx 0.40$, $p^{\text{N}_{\text{C}}^+}_{\text{max}} \approx 0.0024$. These low relative populations are consistent with the fact that the concentration of N$_\text{C}$ is typically much larger than the concentration of N$V$ centers in diamond samples. We note here that when the boron-induced shifts are dropped so that the only shift is that induced by the ether-terminated diamond surface, then $p^{\text{N}_{\text{C}}^+} \approx 0.010$, which is consistent with the relative concentrations of defects reported in previous works~\cite{Broadway2018spat,Kuate2023theor}.

\section{Conclusion}
Ultimately, we have demonstrated that our charge-transfer formalism is able to allow for the investigation of partitioning of the relative populations of defects that may only be or have only been indirectly probed in experiments. Our present treatment also affirms that a simple model incorporating only charge-state transition levels and relative population corrections is sufficient to capture charge-state decay and allows for the determination of local relative concentrations of ionized defects. We anticipate that these results may help inform further experiments to comprehensively capture the charge-state dynamics of NV$^-$ centers in various environments.   
\label{sec:conc}

\section*{Acknowledgments}
We gratefully acknowledge financial support from the College of Engineering and Computer Science at Syracuse University and the use of the Zest High Performance Computing cluster of Syracuse University.

\bibliography{refs_NV}
\end{document}